\documentclass[10pt,conference]{IEEEtran}

\usepackage{graphicx}

\begin{document}

\title{A Fluid Approach for Poisson Wireless Networks}

\makeatletter
\let\thanks\@IEEESAVECMDthanks%
\makeatother

\author{
\authorblockN{
Jean-Marc Kelif$^1$,\thanks{$^1$Jean-Marc Kelif is with Orange Labs, France} 
\thanks{Email: jeanmarc.kelif@orange.com} 
Stephane Senecal$^2$,\thanks{$^2$Stephane Senecal is with Orange Labs, France} 
\thanks{Email: stephane.senecal@orange.com} 
Constant Bridon$^3$,\thanks{$^3$Constant Bridon is with ENS Cachan and Orange Labs, France} 
\thanks{Email: constant.bridon@ens-cachan.fr} 
Marceau Coupechoux$^4$ \thanks{$^4$Marceau Coupechoux is with Telecom ParisTech, France} \thanks{Email: marceau.coupechoux@telecom-paristech.fr}
}}


\maketitle

\begin{abstract}

Among the different models of networks usually considered, the hexagonal network model is the most popular. However, it requires extensive numerical computations. The Poisson network model, for which the base stations (BS) locations form a spatial Poisson process, allows to consider a non constant distance between base stations. Therefore, it may characterize more realistically operational networks. The Fluid network model, for which the interfering BS are replaced by a continuum of infinitesimal interferers, allows to establish closed-form formula for the SINR (Signal on Interference plus Noise Ratio). This model was validated by comparison with an hexagonal network. The two models establish very close results.
In this paper, we show that the Fluid network model can also be used to analyze Poisson networks.
Therefore, the analysis of performance and quality of service becomes very easy, whatever the type of model, by using the analytical expression of the SINR established by considering the fluid model.
\end{abstract}

%

\maketitle

\section{Introduction}
Performance and quality of service (QoS) evaluations of wireless networks can be analyzed by using simulations or analytical models. 
Several QoS parameters (like throughput, outage probability) can be derived from the SINR distribution. Analytical models thus try to derive simple SINR formula in order to quickly evaluate the performance of a cellular network.
Due to the explosion of mobile services demand, the estimation of performance and QoS has to be more and more precise. Therefore, their analysis need tractable and accurate models of networks.

The most popular wireless network model is the hexagonal one: the transmitting base stations constitute a regular infinite hexagonal grid (\cite{Bon03}-\cite{Vit95}-\cite{Lagr05}). 
Although this model seems rather ``reasonable'' for regular deployments of base stations, it is intractable from an analytical point of view. Therefore, it implies extensive numerical computations. Several numerical techniques have been developed to perform such computations. For example, Monte Carlo simulations are widely used in conjunction with this model \cite{Gil91,Ela05} or numerical computations in hexagonal networks \cite{Vit94,Vit95}.

Let us notice that tractability of a wireless network model allows to dramatically reduce the computation time. For example, several optimization problems that can be solved by metaheuristics like Tabu Search \cite{Kam12} or Simulated Annealing \cite{Sam12}, require extensive SINR calculations. Moreover, a network model highlights the parameters of the system, and thus allows to better understand their impact, especially on the quality of service and performance.
	
Another wireless network model consists in the Poisson model: the base stations are randomly distributed on the considered area according to a spatial Poisson process \cite{Bac03}. Although this model is less popular than the hexagonal one, it allows to take into account a more realistic environment since the distances between base stations are not constant. 
Measurements from operational networks have shown that hexagonal model is rather optimistic  and Poisson model is rather pessimistic.

Another model of network is the fluid model \cite{KeA05} \cite{KelCoEurasip10}. This model considers the interfering base stations as a continuum of infinitesimal interferers distributed in space. 
The main interest of this model consists in its tractability, in the possibility to establish closed form formula of the SINR, whatever the location of a UE, and to establish the SINR distribution \cite{KelCoICC09} \cite{KelCoPhycom11}, too. Furthermore, the fluid model 
was shown to be very close to the intractable hexagonal model in terms of evaluation of the SINR.

\underline{Our contribution}: In this article, we show that the distribution of the SINR can be calculated by using the fluid model, whatever the spatial distribution of base stations, regular hexagonal or random Poisson, and whatever the density of BS. Therefore, the determination of the quality of service and performance of a wireless network can be done in a simple way in all these cases.

The organization of the paper is as follows. In Section~\ref{systemmodel}, we present the system model. We introduce the different types of network models used to analyze wireless networks in Section~\ref{modelnetwork}. We show, in Section \ref{PoissonFluid}, that performance and QoS of Poisson model can be fitted by using a fluid model. A conclusion is given in Section \ref{Conclusion}.

\section{System Model} \label{systemmodel}

Let us consider a wireless network. We focus on the downlink transmission part. Our aim is to evaluate the performance and the quality of service of a single user. We consider an access technology in which the radio resources of a base station (BS) are divided in a number of parallel, orthogonal, non-interfering channels (subcarriers), i.e. OFDMA. Therefore, only inter-cell interference is considered, no intra-cell interference. 

\subsection{SINR of a user} 
We consider a single frequency network composed of $N$ base stations, transmitting at power $P$ on each subcarrier. We define $g_i(u)$ the path gain between BS $i$ and user $u$ on a given subcarrier. The SINR $\gamma_u$ of user $u$ served by BS $i$ on the considered subcarrier is given by:
\begin{equation} \label{SINR}
\gamma_u=\frac{P g_i(u) }{
\sum\limits_{j\neq i} P g_{j}(u)  + N_{th}}.
\end{equation}
with $ N_{th}$ the thermal noise on a subcarrier.

We consider a urban environment, where the thermal noise can be neglected. The SINR can then be expressed as:

\begin{equation} \label{SIR}
\gamma_u=\frac{P g_i(u) }{
\sum\limits_{j\neq i} P g_{j}(u) },
\end{equation}

\subsection{Performance and quality of service} 
The knowledge of the SINR allows to calculate the throughput that may be reached by a user. Indeed, considering any subcarrier as an AWGN (Additive White Gaussian Noise) channel, the SINR received by a mobile enables the determination of the spectral efficiency $D_u$ (in bits/s/Hz) by using the Shannon formula: 

\begin{equation} \label{shannon}
D_u= \log_2(1+ \gamma_u). 
\end{equation}

Let us notice that there are alternative approaches like using a modified upper bounded Shannon formula or throughput-SINR tables coming from physical layer simulations. 
Moreover, expressions (\ref{SIR}) and (\ref{shannon}), calculated at any location of the network, allow an evaluation of the CDF (Cumulative Distribution Function) of the SINR (or the throughput). The SINR CDF also provides the outage probability, i.e. the probability that a user cannot be accepted in the network since he cannot have a sufficient throughput.
It is therefore important to develop a method which allows to determine these characteristics with a high accuracy, for a user at any distance $r$ from his serving BS.

Moreover, since the throughput allows to know the quality of service that can be offered to a user, these methods make it possible to determine this characteristic with a high accuracy in a simple way. In particular, the minimum throughput, obtained at cell edge, can be derived. By doing an integration all over the cell range, the average throughput of the cell can be calculated, too.
Dynamical analysis also need the knowledge of the SINR \cite{Rong11} as input.

\section{Wireless Networks Models} \label{modelnetwork}

In this section, we review three cellular network models: the hexagonal one, the Poisson one and the Fluid one. We also recall how to calculate the SINR thanks to the fluid model.

\subsection{Hexagonal Network}
\begin{figure}[!h]
\centering
\includegraphics[width=0.7\linewidth]{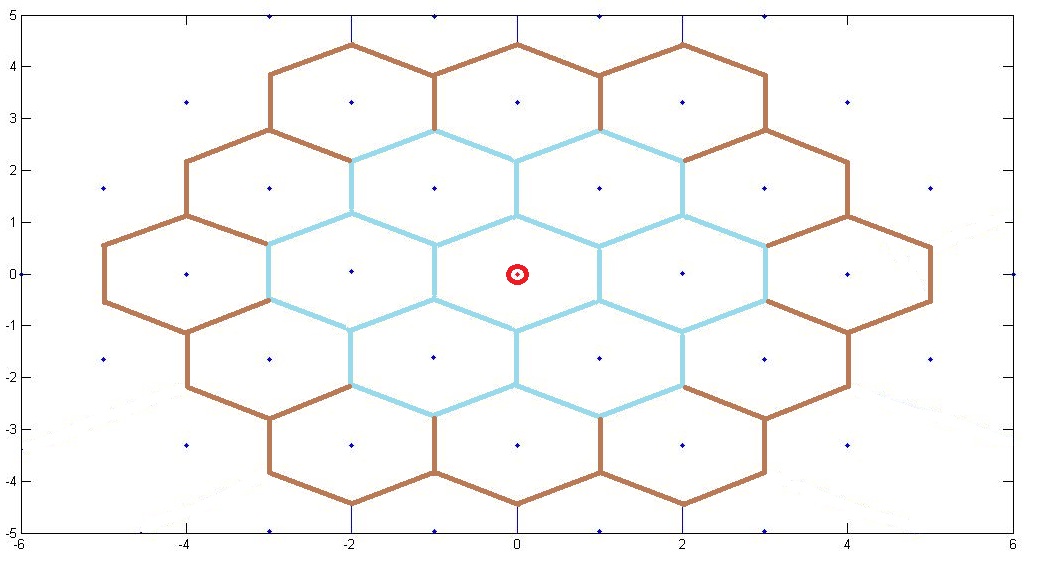}
\caption{Hexagonal Network} 
\label{hexagonalnetwork}
\end{figure}

In a hexagonal configuration, the base stations are regularly distributed in the area (Fig. \ref{hexagonalnetwork}). The model is characterized by a single parameter, which is the Inter Site Distance (ISD). 
Since all the zones covered by any BS are equivalent, it is sufficient to analyze any unique cell in the whole system.

\subsection{Poisson Network}
\begin{figure}[!h]
\centering
\includegraphics[height=5cm]{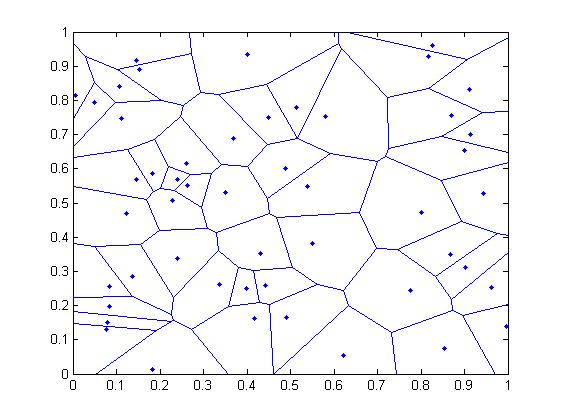}
\caption{Poisson Network} 
\label{Poissonnetwork}
\end{figure}

In a real network, the inter site distance is variable. The Poisson model network, characterized by the density of BS, allows to take it into account by considering a Poisson distribution of base stations in a given area (Fig. \ref{Poissonnetwork}). In this configuration, the cells of the network form a Vorono\"i diagram. Therefore, it becomes necessary to analyze a wide zone, with a great number of base stations, to determine the statistical characteristics of the network in terms of performance and quality of service.

\subsection{Fluid Network}
\begin{figure}[!h]
\centering
\includegraphics[height=5cm]{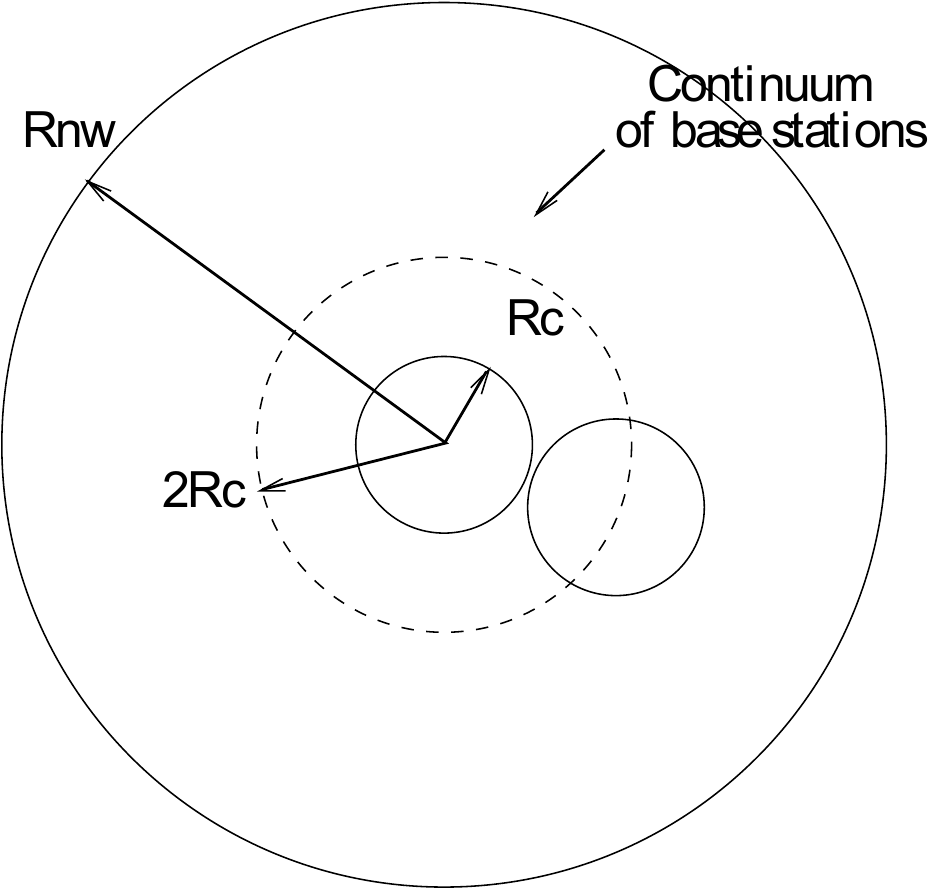}
\caption{Fluid model: Network and cell of interest} 
\label{Fluidnetwork}
\end{figure}

The fluid model consists in replacing a given fixed finite number of transmitters by an equivalent continuous density of transmitters. Given an inter site distance $2R_c$, 
interferers are characterized by a  density $\rho_{BS}$
of BS starting at a distance of $2R_c$ from a BS (covering a zone of radius $R_c$), as illustrated on Fig. \ref{Fluidnetwork} ($R_{nw}$
is the size of the network).

\subsection{Calculation of the SINR}

We consider a path gain $g_{j}(u) = K r_j^{-\eta}(u)$, where  $K$ is a constant, $r_j(u)$ is the distance between user $u$ and BS $j$ and $\eta$ the path loss exponent. Let us consider that a BS (eNode-B) transmits at power $P$ on each subcarrier. Denoting r = $r_i$, we can express (\ref{SIR}) as (dropping u):

\begin{equation} \label{SIRdist}
\gamma=\frac{r^{-\eta} }{
\sum\limits_{j\neq i} r_j^{-\eta}},
\end{equation}

By considering expression  (\ref{SIRdist}), we can see that the calculation of the SINR of a user depends on its location, on the location of its serving BS, and on the location of each interfering BS. 

Considering the fluid model, the SINR only depends on the distance $r$ of the user to its serving base station \cite{Mas11}: 

\begin{equation}
\gamma(r)= \frac{\eta -2}{2\pi \rho_{BS}} \frac{r^{-\eta}}{(2R_c-r)^{2-\eta}}  \label{SIRfluid}
\end{equation}

If $\rho_{BS}$ is known, this analytical model allows to significantly simplify the calculation of the SINR: the only required variable is the distance of the mobile to its serving BS. This model has been proven to be reliable and close to the reality for homogeneous networks \cite{KeA05} \cite{KelCoEurasip10}, as well as for heterogeneous networks \cite{KeSene12}.

The cell edge throughput can particularly be calculated by setting $r$ = $R_c$ in (\ref{SIRfluid}). Therefore, the minimum performance and quality of service offered to UE is evaluated in a simple way. Moreover, a simple integration over the cell range allows to calculate the average throughput of the cell.

These results on the fluid model are valid for a constant inter site distance 2$R_c$, and the purpose of the next section will be to find a correspondence between a completely stochastic network and the fluid model.

\section{Fluid Model of Poisson Network} \label{PoissonFluid}
In a real network, as mentioned in the introduction, the inter site distance is evidently variable. However, the fluid model relies on the hypothesis of a regular network. This section shows that it is possible to adapt the fluid model in order to find an equivalence between a Poisson random network and a regular network associated to the fluid model in terms of CDF of the SINR.


The first studied network is a Poisson network. We consider a given area of surface $S_A$. 

\subsection{Base stations distribution}
In order to place the base stations, we set the \textit{expected half inter site distance} as $R_c$. This hypothesis fixes the density of base stations $\rho_{BS}$. The values of $\rho_{BS}$ and of the studied surface $S_A$ give the Poissonian characteristic of the network: the number of BS is drawn according to a Poisson distribution of parameter $\rho_{BS} S_A$. The surface is chosen to obtain in average 50 stations in the area. It allows to have a significant number of cells, representative of a realistic zone covered by BS, and a significant number of interfering BS for the computation of the SINR. Those BS are then placed in the network, with no pairwise constraint (Fig. \ref{Poissonnetwork}): distances between neighboring base stations may be very low.

\subsection{Poisson network SINR computation}
The users are uniformly distributed on the whole area $S_A$. Then, the SINR of a UE is computed from its definition: the best serving BS gives the power of the received signal, and all the other stations generate interference. Several Monte Carlo simulations are run. At each run, the number and locations of the BS change, whereas the set of studied points (UE) is fixed. As a result, for the set of studied points, we obtain the corresponding SINR with different configurations of BS. Therefore, it becomes easy to compute the CDF of the SINR received by UE in this zone. Considering a toroidal shape of the network allows to consider it as virtually infinite with no ``edge effect'' for the computation of the SINR.  

\subsection{Fluid network SINR calculation}
As a comparison of this result, we calculate the SINR in a cell of radius $R_c$ with the fluid model by using (\ref{SIRfluid}). Let us recall that the fluid model has been proven to be a good approximation for the computation of the SINR for hexagonal networks \cite{KelCoEurasip10}.

\subsection{Comparison of the SINR CDF for the Poisson and Fluid models}
\begin{figure}[htbp]
\begin{center}
\includegraphics[width=1\linewidth]{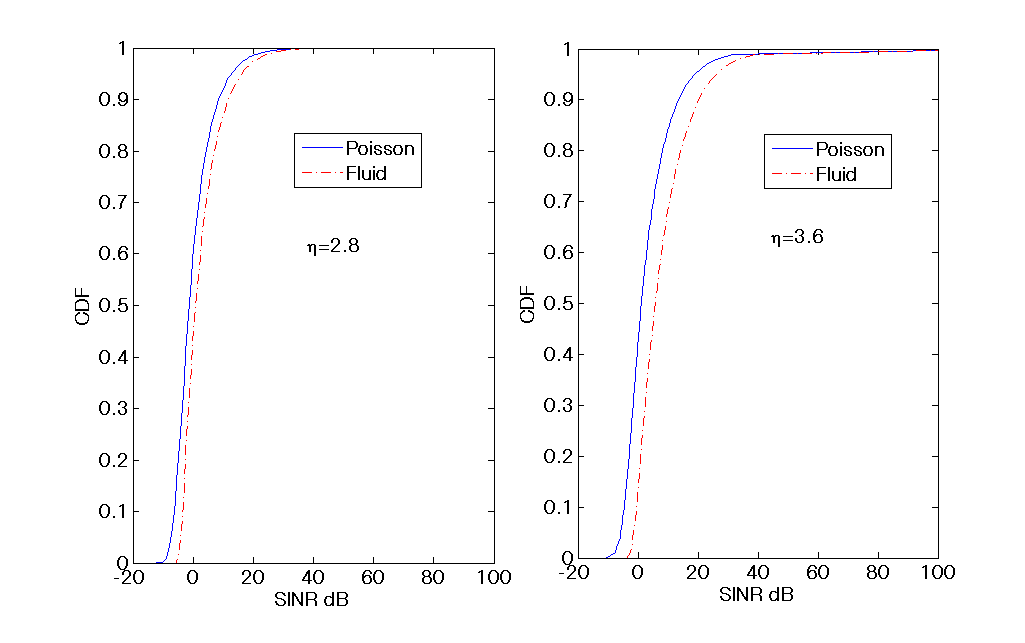}
\caption{CDF of the SINR with $\eta$ = 2.8 (left) and $\eta$ = 3.6 (right), for a Poisson model network and a Fluid model network.}
\label{sinrcdfnonfitted2836}
\end{center}
\end{figure}

Since the CDF of the SINR characterizes the performance and the quality of service of wireless systems, we establish the CDF of the SINR, considering the Fluid network and the Poisson network for a wide range of values of $\eta$, comprised between 2.2 and 4.2. Fig. \ref{sinrcdfnonfitted2836} shows an example of the curves established for $\eta$ = 2.8 and 3.6.  We first notice that a Poisson network gives always lower values of SINR than a regular hexagonal network: let us recall that the fluid model was validated by comparison to a hexagonal network. The difference between the two curves is about 2 dB for\\
$\eta$ = 2.8 and 4 dB for $\eta$ = 3.6. 
These results mean that the Poisson network is rather ``pessimistic'' and the Fluid network ``optimistic''. It can be explained by the fact that there is a probability to have interfering base stations very close to a user in a Poisson network. 
Another noticeable aspect is the difference of those two curves along the abscissa axis. In fact, observations point out that the CDF of the SINR of the Poisson network is similar to the CDF of the SINR established by using the Fluid network model, translated along the abscissa axis. Moreover, this translation seems to depend on the value of the propagation parameter $\eta$. The simulations take into account 11 different values of $\eta$ with an increment of $0.2$, ranging from 2.2 to 4.2 (usual range for the path-loss exponent is comprised between 2.8 and 3.6).

\subsection{SINR CDF Fitting}
As mentioned, the difference between the two SINR CDF curves along the abscissa axis seems to be constant, and increases with $\eta$. Therefore, we apply a curve fitting with a polynomial approximation. In this aim, we compute the mean value of that difference for each $\eta$. Then, the comparison of this value with a polynome of degree 1 in $\eta$ is sufficient to achieve an excellent approximation. The result turns out to be interesting: the linear expression in $\eta$ fits the curve with the same coefficients, whatever the value of $\eta$. The SINR calculated is corrected by the linear expression, and in this case, we establish the formula:
\begin{equation}
SINR^{fitted}_{Fluid} = SINR_{Fluid}-(a \eta + b)
\label{fitting}
\end{equation}
which yields
\begin{equation}
CDF_{Poisson} \approx CDF^{fitted}_{Fluid}
\label{cdffitting}
\end{equation}
where $a=$ 3 and $b=$ -6.

\begin{figure}[htbp]
\begin{center}
\includegraphics[width=1\linewidth]{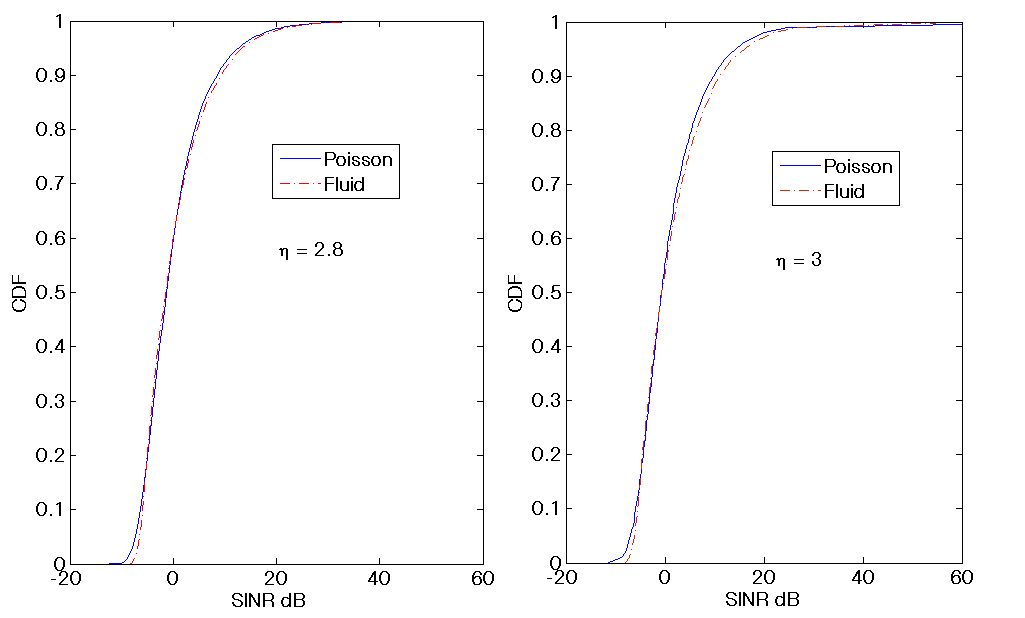}
\caption{CDF of the fitted SINR with $\eta$ = 2.8 (left) and $\eta$ = 3 (right), for a Poisson model network and a Fluid model network.}
\label{sinrcdffitted2830}
\end{center}
\end{figure}

\begin{figure}[htbp]
\begin{center}
\includegraphics[width=1\linewidth]{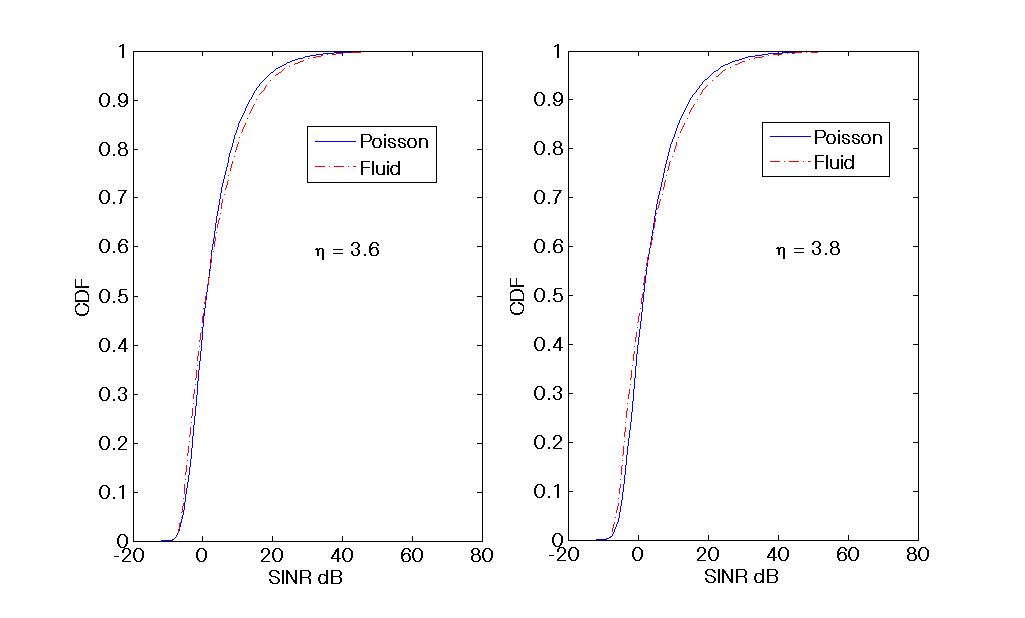}
\caption{CDF of the fitted SINR with $\eta$ = 3.6 (left) and $\eta$ = 3.8 (right), for a Poisson model network and a Fluid model network.}
\label{sinrcdffitted3638}
\end{center}
\end{figure}

As observed on some examples (Fig. \ref{sinrcdffitted2830} and \ref{sinrcdffitted3638}), for $\eta$ = 2.8, 3, 3.6, 3.8, the CDF of the SINR established by the \textit{fitted Fluid model} and by the \textit{Poisson model} are very close: the differences between them are less than 0.4 dB (let us notice that, for an outage probability of 5\%, this difference may reach 0.9 dB for some values of $\eta$).

\subsection{Impact of the density of BS} 
\begin{figure}[htbp]
\begin{center}
\includegraphics[width=1\linewidth]{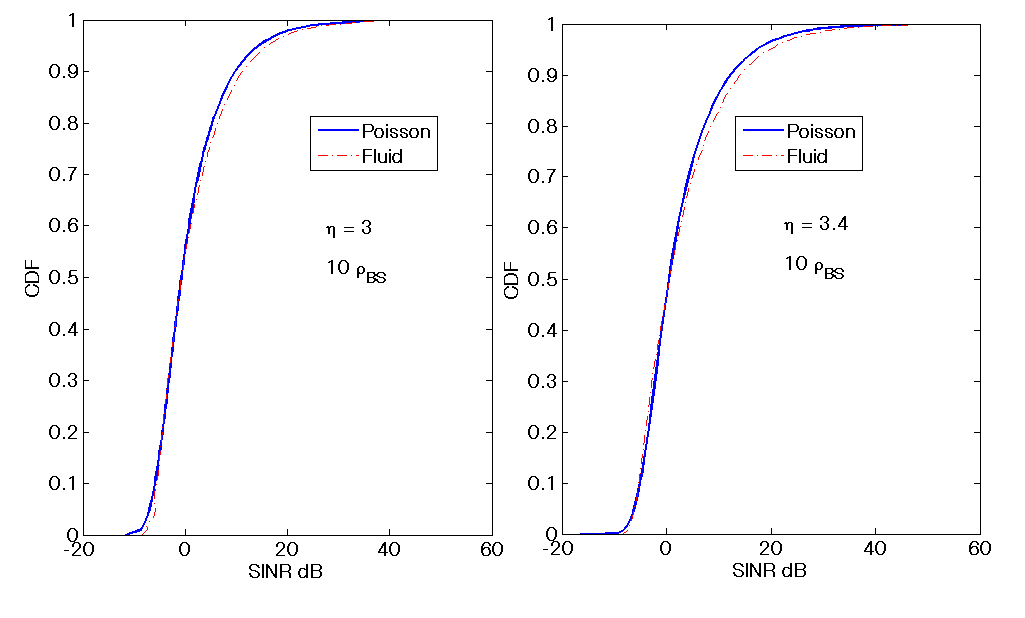}
\caption{CDF of the fitted SINR with $\eta$ = 3 (left) and $\eta$ = 3.4 (right), for a Poisson model network and a Fluid model network, for a density of BS equals to 10 $\rho_{BS}$.}
\label{cdf10rho}
\end{center}
\end{figure}

\begin{figure}[htbp]
\begin{center}
\includegraphics[width=1\linewidth]{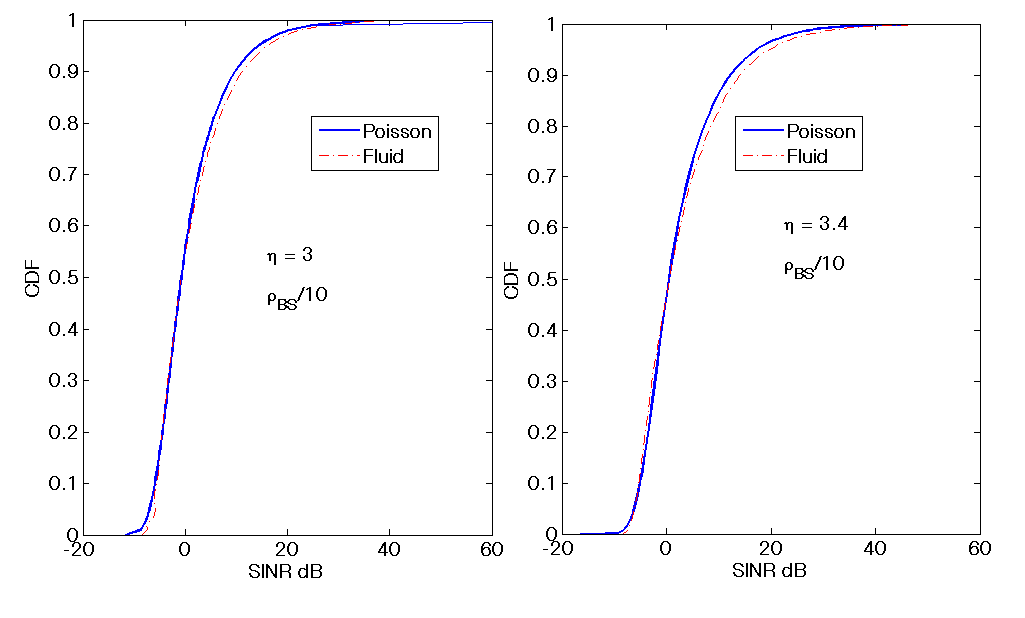}
\caption{CDF of the fitted SINR with $\eta$ = 3 (left) and $\eta$ = 3.4 (right), for a Poisson model network and a Fluid model network, for a density of BS equals to $\rho_{BS}/10$.}
\label{cdfrhosur10}
\end{center}
\end{figure}

It is interesting to observe that the fitting does not depend on the density of BS. Fig. \ref{cdf10rho} and \ref{cdfrhosur10} show that when the density is multiplied by a factor 10 or divided by a factor 10, the CDF are identical, and the fitting is the same.
As a consequence, these results can be used whatever the density of the network, i.e. whatever the intersites distance between base stations. 
It is easy to understand this property when we realize that, since $\rho_{BS}=\frac{\sqrt{3}}{6 R_c^2}$, the expression of the SINR (\ref{SIRfluid}) given by the fluid model, can be expressed as:

\begin{equation}
\gamma(x)= \frac{6}{\sqrt{3}}\frac{\eta -2}{2\pi} \frac{x^{-\eta}}{(2-x)^{2-\eta}}  \label{gammax}
\end{equation}

where $x= \frac{r}{R_c}$ represents the relative distance of a UE to its serving base station. This expression does not explicitely depend on the ISD (distance between neighboring BS). This expression holds whatever the ISD. Moreover, it does not depend on the density of BS.

\subsection{Correlation coefficient}
We compare the correlation coefficient (\ref{correlation}) computed between the CDF curves of the fitted fluid model and of the Poisson model. We recall that, given $X$ and $Y$, two samples of length $n$ with respective means $\bar{X}$ and $\bar{Y}$, the correlation coefficient $\zeta$ is given by:
\begin{equation}
\zeta=\frac{\sum_{i=1}^{n}(X_i-\bar{X})(Y_i-\bar{Y})}{\sqrt{\sum_{i=1}^{n}(X_i-\bar{X})^2}\sqrt{\sum_{i=1}^{n}(Y_i-\bar{Y})^2}}
\label{correlation}
\end{equation}
This comparison is synthetized in Table \ref{coefcorrelation} illustrating that these correlation coefficients are excellent. The linear approximation becomes less accurate for the minimum and the maximum values of $\eta$: 2.4 and 4. Indeed, when this coefficient is not better than 0.99, the correlation is usually considered as not good.

%
%

\begin{table}[htbp]
\begin{center}
\begin{tabular}{|c|c|} 
\hline $\eta$ & Correlation coefficient $\zeta$ \\
\hline  2.2 & 0.9755   \\
\hline  2.4 & 0.9961   \\
\hline  2.6 & 0.99909   \\
\hline  2.8 & 0.99966   \\
\hline  3   & 0.99952   \\
\hline  3.2 & 0.99941   \\
\hline  3.4 & 0.99934   \\
\hline  3.6 & 0.99897   \\
\hline  3.8 & 0.99861   \\
\hline  4   & 0.99824   \\
\hline  4.2 & 0.99683   \\
\hline 
\end{tabular} 
\end{center} 
\caption{Correlation coefficient for the comparison of the CDF of the SINR from the fitted Fluid network and the Poisson network with respect to $\eta$}
\label{coefcorrelation} 
\end{table}%

\subsection{Model limitations} The fluid model with a simple linear fitting in $\eta$ with known parameters gives the same CDF of the SINR as a Poisson network computing the SINR by considering (\ref{SIRfluid}) and (\ref{fitting}). The fluid model is thus also reliable with random networks. For values of $\eta$ lower than 2.6 or higher than 3.8 (i.e. out of the range of usual values of $\eta$), it can be observed that the proposed fitting is less accurate. 

In conclusion, in this validity field, we can say that there is an equivalence between the Fluid network and a Poisson network, in terms of CDF of SINR, therefore in terms of outage probability and throughput. The fluid model can thus be used to evaluate the performance and quality of service of any kind of network, hexagonal and Poisson.

%
%

%
%
%


\section{Conclusion}\label{Conclusion}
In this paper, we established that the Fluid network model can be used to analyze Poisson networks as well as hexagonal networks. We have shown  that the CDF of the SINR, calculated by the analytical expression of the fluid model network and for the Poisson model network are very close, when a simple fitting with a linear function of the propagation parameter $\eta$ is applied. We have also shown that this is true regardless of the density of base stations, since the analytical expression is insensitive to density. Therefore, the analysis of performance, outage probability, throughput (which are important key performance indicators for quality of service) becomes very easy, whatever the type of model, by using the analytical expression of the SINR established by the Fluid network model.

\section*{Acknowledgment}
This work was partially supported by the Seventh Framework Programme for Research of the European Commission under grant number
HARP-318489 and by the ANR France NETLEARN project.


\end{document}